\documentstyle[12pt]{article}
\begin{document}

\title{Algebra of Constraints and Solutions of Quantum Gravity}
\author{A.\ B\l{}aut\thanks{e-mail address
ablaut@ift.uni.wroc.pl} and J.\ Kowalski--Glikman\thanks{e-mail 
address
jurekk@ift.uni.wroc.pl}\\
Institute for Theoretical Physics\\
University of Wroc\l{}aw\\
Pl.\ Maxa Borna 9\\
Pl--50-204 Wroc\l{}aw, Poland}
\maketitle

\begin{abstract}
We construct the regularized Wheeler--De Witt operator demanding that 
the
algebra of constraints of quantum gravity is anomaly free. We find 
that
for only a small subset of all wavefunctions being integrals of scalar
densities this condition can be satisfied. It turns out that the
resulting operator is much simpler than the one used in \cite{JK} to 
find
exact solutions of Wheeler--De Witt equation. We proceed to finding
exact solutions of quantum gravity and we discuss their interpretation
making use of the quantum potential approach to quantum theory.
\end{abstract}
\clearpage

\section{Introduction}
One of the most outstanding problems of modern theoretical physics is
the construction of quantum theory of gravity \cite{reviews}. Indeed, 
it have been
claimed many times that various unsolved problems like the 
cosmological
constant problem, the problem of origin of the universe, the problem 
of
black holes radiation will find their ultimate solution once this 
theory
is finally constructed and properly understood. Some \cite{Emperor},
claim that the theory of quantum gravity will also shed some light on 
the
fundamental problems of quantum mechanics and even on the origin of 
mind.
These all prospects are very exciting indeed, however the sad fact
remain that the shapes of the future theory are still very obscured.

Nowadays there are two major ways of approaching the problem of
quantum gravity. The first one is associated with the broad term
``superstrings''. In this approach the starting point is a
two-dimensional quantum field theory which yields quantum gravity as 
one
of resulting low-energy  effective theories. It is clear that in
superstrings, as in other, less
developed approaches in whose gravity appears as an effective theory, 
it
does not make sense to try to ``quantise'' classical gravity.

In the
canonical approach one does something opposite: the idea is to pick up
some structures which appear already at the classical level and then
promote them to define the quantum theory. In both the standard
canonical approach in metric representation, which we will follow 
here, and
in the approach based
on loop variables \cite{loop}, these fundamental structures are 
constraints
of the  classical canonical formalism reflecting the symmetries of the
theory  and their algebra. There are good
reasons for such an approach. The equivalence principle is the main
physical principle behind the classical theory of gravity; this
principle leads to the general coordinate invariance and selects the
Einstein--Hilbert action as the simplest possible one.

Another building block of the quantum theory is the quantisation
procedure. Here one encounters the problem as to if a generalization 
of
the standard Dirac procedure of quantisation of gauge theories is
necessary. This would be the case if one shows that the standard 
approach
is not capable of producing any interesting results. It is not 
excluded
that this may be eventually the result of possible failure of
investigations using standard techniques, however, in our opinion, at 
the moment there
is no reason to modify the basic principles of quantum theory.

Our starting point consists therefore of
\begin{itemize}
\item[(i)] The classical constraints of Einstein's gravity: the
diffeomorphism constraint generating diffeomorphism of the spatial
three-surface ``of constant time''
\begin{equation}
{\cal D}_a=\nabla_b\, \pi^{ab}
\end{equation}
and the hamiltonian constraint generating ``pushes in time 
direction'':
\begin{equation}
{\cal H} = \kappa^2 G_{abcd}\pi^{ab}\pi^{cd} - \frac1{\kappa^2}\sqrt h
(R +2\Lambda)
\end{equation}
In the formulas above $\pi^{ab}$ are momenta associated with the
three-metric $h_{ab}$,
$$
G_{abcd}=\frac1{2\sqrt
h}\left(h_{ac}h_{bd}+h_{ad}h_{bc}-h_{ab}h_{cd}\right)
$$
is the Wheeler--De Witt metric, $R$ is the three-dimensional
curvature scalar, $\kappa$ is the gravitational constant, and 
$\Lambda$ the
cosmological constant. The constraints satisfy the following algebra
\begin{equation}
[{\cal D}, {\cal D}] \sim {\cal D},\label{difdif}
\end{equation}
\begin{equation}
[{\cal D}, {\cal H}] \sim {\cal H},\label{difham}
\end{equation}
\begin{equation}
[{\cal H}, {\cal H}] \sim {\cal D}. \label{hamham}
\end{equation}
\item[(ii)] The rules of quantisation given by the metric 
representation
of the canonical commutational relations
$$
\left[\pi^{ab}(x),h_{cd}(y)\right]=-
i\delta^{(a}_c\delta^{b)}_d\delta(x,y),
$$
$$
\pi^{ab}(x)=-i\frac{\delta}{\delta h_{ab}(x)}.
$$
\end{itemize}

Sadly, in the canonical approach, the points (i) and (ii) above
encompass the whole of the input in our disposal in construction of 
the
quantum theory. In particular, we do not know what is the correct
physical inner product, and thus we do not know if the relevant 
operators
are hermitean or not. Besides, we do not even know if we should demand
these operators to be hermitean: the hamiltonian annihilates the
physical states (the famous time
problem \cite{ishamtime}) and thus unitary evolution does not play the
privileged role anymore. It follows that we cannot distinguish
``relevant''
wave functions by demanding that they are normalizable, as in the case
of quantum mechanics, in fact, since the probabilistic interpretation 
of
the ``wavefunction of the universe'' is doubtful, it is not clear if 
the
norm of this wavefunction is to be 1.

In the recent paper \cite{JK} a class of exact solutions of the
Wheeler--De Witt equation was found. In that paper we used the heat
kernel to regularize the hamiltonian operator and inserted the
particular ordering. The question arises what is the level of
arbitrariness in this construction. In other words, could we construct
other (possibly simpler) regularized hamiltonian operators and what 
would be their
properties? This question is the subject of the present paper.

It is clear from the discussion above that the only principle, we can
base our construction on is the principle that the algebra of
constraints is to be anomaly--free, that is, the corresponding algebra
of commutators of quantum constraints is weakly identical with the
classical one. This means that the structure of the
Poisson bracket
algebra (\ref{difdif}--\ref{hamham}) is to be preserved, in the sense
which will be explained below, on the quantum level. The following
section is devoted to the analysis of this problem.
In section 3 we investigate solutions of the resulting equations,
and in section 4 we seek interpretation of the wavefunctions making 
use
of the quantum potential approach to quantum mechanics. In the final
section we draw our conclusions and describe the open problems.

\section{The commutator algebra and construction of regularized
operators}

As we explained in Introduction, our starting point in construction of
the quantum hamiltonian operator (the Wheeler--De Witt operator) is 
the
algebra (\ref{difdif}--\ref{hamham}) and we demand that the same 
algebra
holds on the quantum level. At this point we encounter immediately the
problem, well known from the investigations of anomalies in quantum 
field
theories, that the sole algebra of regularized operators is 
meaningless
unless the space of states on which these operators act is defined 
{\em
a priori}\footnote{Here our approach differs from the one used in, for
example \cite{Maeda}, where the authors choose to analyze the algebra 
of
quantum constraints without defining the space of states.}. This 
follows
from the fact that the transition from regularised to renormalised
action of an operator depends crucially on what particular state this
operator acts (see below.) We will chose
our starting space of states to be the space of
integrals over compact three-space $M$ of scalar densities like ${\cal
V}=\int_M\sqrt h$, ${\cal
R}=\int_M\sqrt h R$, etc.;
\begin{equation}
\Psi = \Psi({\cal V}, {\cal R}, \ldots).
\end{equation}

We choose the following representation of the diffeomorphism
constraint
$$
{\cal D}_a(x)=-i\nabla_b^{x} \frac{\delta}{\delta h_{ab}(x)},
$$
where we employed the notation $\nabla_b^{x}$ meaning that the 
covariant
derivative acts at the point $x$. Then we see that diffeomorphism
constraint annihilates all the states and the commutator relation
(\ref{difdif}) is identically satisfied. Moreover we see that the 
relation
(\ref{difham}) reduces to the formal relation
\begin{equation}
{\cal D}({\cal H}\Psi)\sim {\cal H}\Psi.\label{difham1}
\end{equation}

Now we must turn to the heart of the problem, the construction of the
Wheeler--De Witt operator. It is well known that second functional
derivative acting at the same point on a local functional
produces divergent result. We
deal with this problem by making the point split in the kinetic term, 
to
wit
$$
G_{abcd}(x)\pi^{ab}(x)\pi^{cd}(x) \Longrightarrow
\int\, dx'\, K_{abcd}(x,x';t)
\frac{\delta}{\delta h_{ab}(x)}\frac{\delta}{\delta h_{cd}(x')},
$$
where $K_{abcd}(x,x';t)$ satisfies
$$
\lim_{t\rightarrow 0^+}K_{abcd}(x,x';t)=\delta(x,x').
$$
By virtue of the correspondence principle, we take
\begin{equation}
K_{abcd}(x,x';t)=G_{abcd}(x')\triangle(x,x';t)\left(1 + K(x,t)\right),
\end{equation}
where
$$
\triangle(x,x';t)=\frac{\exp\left(-\frac1{4t}
h_{ab}(x-x')^a(x-x')^b\right)}{4\pi t^{3/2}}
$$
and
$K(x,t)$ is analytic in $t$.      

Next we must resolve the ordering ambiguity in the operator ${\cal 
H}$. To
this end we add the new term $L_{ab}(x)\frac{\delta}{\delta 
h_{ab}(x)}$,
where $L_{ab}$ is a tensor to be derived along with $K(x,t)$. Thus the
final form of the Wheeler--De Witt operator is\footnote{In the paper
\cite{JK} we took $\tilde K_{abcd}(x,x')=G_{abcd}(x) \tilde K(x,x')$,
where $\tilde
K$ is a heat kernel, and $L_{ab}$ was taken to be the functional
derivative of $\tilde K_{abcd}$ with respect to $h_{cd}$.}
$$
{\cal H}(x)=\kappa^2\int\, dx'\, K_{abcd}(x,x';t)
\frac{\delta}{\delta h_{ab}(x)}\frac{\delta}{\delta h_{cd}(x')} +
$$
\begin{equation}
+ L_{ab}(x)\frac{\delta}{\delta h_{ab}(x)}
+ \frac1{\kappa^2}\sqrt h
(R +2\Lambda).\label{qham}
\end{equation}

To set the stage, we still need to define the action of operators on
states. To this end we must discuss the issue of regularization
and renormalization. The operator (\ref{qham}) acting on a state
(defined as an integral of a scalar density) produces, in general, 
terms
with arbitrary (positive and negative) powers of $t$. This provides 
the
regularized version of the operator since all the terms are finite, 
and
singularities of the form $\delta(0)$ are traded for terms which are
singular for $t\rightarrow0$. Observe that, as noted above, the 
singular
part of the action of the operator on a state depends on this state. 
To
renormalize, we follow the procedure
proposed by Mansfield \cite{Mansfield} which result in the following:
the terms with positive powers of $t$ are dropped, and the singular
terms of the form $t^{-k/2}$ are replaced by the renormalization
coefficients $\rho^k$. This procedure provides us with the finite 
action of
the Wheeler--De Witt operator on any state.

Now we can turn to the interpretation of equation (\ref{difham1}). We
understand it in the following way. An operator acts on a state and
after renormalization gives another state depending on renormalization
constants. On this resulting state the second operator acts. Thus the
formal relation (7) is defined to mean
\begin{equation}
{\cal D}\,({\cal H}\Psi)_{ren}\sim ({\cal 
H}\Psi)_{ren},\label{difhamf}
\end{equation}
and, similarly, for the hamiltonian-- hamiltonian commutator
\begin{equation}
\left({\cal H}[N]\,\,({\cal H}[M]\Psi)_{ren}\right)_{ren}
- \left({\cal H}[M]\,\,({\cal 
H}[N]\Psi)_{ren}\right)_{ren}=0\label{hamhamf}
\end{equation}
since $\Psi$ is diffeomorphism invariant. In the formula above we used
the smeared form of the Wheeler--De Witt operator
$$
{\cal H}[M]=\int dx\, M(x){\cal H}(x).
$$

Let us turn back to equation (\ref{difhamf}). Since the action of
diffeomorphism is standard, it suffices to check that $({\cal
H}\Psi)_{ren}$ is a scalar density. But this is clearly the case: the
first functional derivative acting on a state produces a tensor 
density
${\sf T}^{ab}(x')$. After acting by the second derivative and
contracting indices, we obtain the terms of the form
$$
{\sf T}_0(x')\delta(x',x) + {\sf T}_1(x')\circ \nabla^{x'}\circ 
\nabla^{x'}
\delta(x',x) + {\sf T}_2(x')\circ \nabla^{x'}\circ \nabla^{x'}\circ
\nabla^{x'}\circ \nabla^{x'}\delta(x',x) + \ldots
$$
where $\circ$ denotes various indices contractions, and ${\sf T}_n$ 
are
tensor densities. These terms are multiplied by $\triangle(x,x';t)$
and integrated over $x'$. Now we integrate by parts which results in
replacing
covariant derivatives acting on $K$ with appropriate powers of $t^{-
1}$
multiplied by some coefficients. After renormalization we obtain a
scalar density as required. The action of the $L$ term clearly gives 
the
same result. Thus
\newline

{\em For the states being integrals of scalar densities there is no
anomaly in the diffeomorphism --- hamiltonian commutator}
\newline

This result is quite important because the anomaly in the string 
theory
appears in the diffeomorphism --- hamiltonian commutator.
\newline

Now we turn to the most complicated problem, the hamiltonian
--- hamiltonian commutator (\ref{hamhamf}). Our goal will be to find 
the
maximal space of states together with conditions defining coefficients
$K$ and $L_{ab}$ of the Wheeler--De Witt operator. Our approach is 
based on
the following
\newline

{\sc Claim}. {\em If $({\cal H}\Psi)_{ren}$ contains terms which 
contain
four or more derivatives of the metric like $R^2$,
$R_{ab}R^{ab}$ etc., then (\ref{hamhamf}) cannot be
satisfied.}
\newline

We have checked this claim for terms proportional to square of
three-curvature; it is clear from this computation that the claim 
holds
for higher order terms as well unless there are some miraculous
cancellations. We leave it as an open problem to check if the claim
above is generally valid.

Let us start with the
simplest state $\Psi=1$. Then the action of the first smeared operator
gives
$$
({\cal H}[M]\Psi)_{ren} = \int dx\, \sqrt h(x) M(x)
(R(x) +2\Lambda).
$$
After rather tedious computation one finds in the commutator the
term proportional to $N\nabla^a M-M\nabla^a M$ which must
vanish, to wit
\begin{equation}
\rho^{(1)}\frac32\nabla_a K - \frac1{\kappa^2}\left(\nabla_a L +
\nabla_b L_a^b\right)
=0,\label{coeff}
\end{equation}
where $L= h_{ab}L^{ab}$.

Turning to the states depending of ${\cal V}=\int_M d^3x\, \sqrt h$, 
we
find, taking the Claim above into account that $K$ and $L_{ab}$ can 
only
contain terms at most linear in Ricci tensor. Given that, there is no
anomaly. Thus we take
$$
L_{ab}=\frac1{\kappa}\alpha h_{ab} + \kappa(\beta h_{ab} R +\gamma 
R_{ab})
$$
where in the first term we included the gravitational constant for
dimensional resons.

All states depending on integrals of scalars constructed from powers 
of
curvature tensor will necessarily lead to terms excluded by virtue of
the Claim. This means that not all states of this form will lead to 
the
anomaly-free algebra: the wavefunction will have to satisfy equations
guaranteeing that such terms are absent. These equations, for the case
of states depending on ${\cal R}=\int_M d^3x\, \sqrt h R$ will
further restrict the form of the regularized Wheeler--De Witt 
operator.

Now we turn to the wavefunction $\Psi=\Psi({\cal R})$. As we argued
above, in the action of the Wheeler--De Witt operator on this state 
all
terms with four derivatives must vanish. These terms are
\begin{equation}
\kappa^2\frac{\partial^2 \Psi}{\partial{\cal R}^2}\left( R_{ab}R^{ab} 
-\frac38
R^2\right)   +
\kappa\frac{\partial \Psi}{\partial{\cal R}}\left(-\gamma 
R_{ab}R^{ab} +
\frac12(\gamma + \beta)
R^2\right) =0,\label{R2}
\end{equation}
from which we obtain conditions on the coefficients
\begin{equation}
\beta = -\frac14\gamma,\label{solcoeff}
\end{equation}
and from (\ref{R2}) we find that
$\Psi({\cal R})$ must be of the form
$$
\Psi({\cal R})= \exp\left(\frac{\gamma}{\kappa}{\cal R}\right).
$$
Thus $L_{ab} = \frac{\kappa}\alpha h_{ab} + \kappa\gamma(R_{ab}-
\frac14 h_{ab} R)$. But
then it follows from (\ref{coeff}) that $K(x,t)$ must be constant. It 
is
possible, in principle, to construct $K$ from global integrals like
${\cal V}$ and/or ${\cal R}$, for example $K=t\frac{{\cal R}}{{\cal
V}}$, however
  we will not pursue this (interesting) possiblity here.

Thus the final form of the regularised Wheeler--De  Witt operator is
$$
{\cal H}(x)=\kappa^2\int\, dx'\, G_{abcd}(x')\triangle(x,x';t)
\frac{\delta}{\delta h_{ab}(x)}\frac{\delta}{\delta h_{cd}(x')} +
$$
\begin{equation}
+ \left(\frac1{\kappa}\alpha h_{ab}+ \kappa\gamma(-\frac14h_{ab}R + 
R_{ab})\right)(x)
\frac{\delta}{\delta h_{ab}(x)}+
 \frac1{\kappa^2}\sqrt h
(R +2\Lambda).\label{qhamf}
\end{equation}

The formula (\ref{qhamf}) completes our construction of the
Wheeler--De Witt operator. As compared to the choice made in the 
paper
\cite{JK}, where we used the heat kernel and $L_{ab}$ was its 
functional
derivative, here we gained much more freedom in the form of two
independent constants. In particular, we can make the constants 
$\alpha$
and $\gamma$ complex. This fact
is very important in view of the quantum potential interpretation of 
our
results (see Section 4), where it turns out that only complex
wavefunctions lead to time-evolving universes.
\newline

\section{Solutions}

From the previous section we know that the most general form of the
Wheeler- De Witt operator satisfying our criteria is given by 
equation
(\ref{qhamf}), with coefficients $\alpha$ and $\gamma$ being still 
not fixed.
Now, employing this operator, we will try to find a class of
solutions of the
Wheeler--De Witt equation. It should be stressed at this point that 
we
regard the existence of a maximal possible space of solutions as an 
ultimate
condition fixing the operator completely. Thus our goal is twofold: 
to
find solutions and to fix the operator as to allow for the maximal
possible number of them.

We will consider only the states of the form $\Psi = \Psi({\cal V},
{\cal R})$. Since we have already taken care of the terms 
proportional
to squares of Ricci tensor in (\ref{solcoeff}), we have two equations
for multipliers of $\sqrt h R(x)$ and $\sqrt h$. They read,
\newline

for $\sqrt h R$:
\begin{equation}
-\frac14\kappa^2\frac{\partial^2\Psi}{\partial{\cal V}\partial{\cal 
R}}+
\frac78\kappa^2\rho^{(3)}\frac{\partial\Psi}{\partial{\cal R}} +
\frac18\kappa\gamma\frac{\partial\Psi}{\partial{\cal V}} +
\frac1{2\kappa}\alpha\frac{\partial\Psi}{\partial{\cal R}} +
\frac1{\kappa^2}\Psi=0,\label{R}
\end{equation}
\vspace{12pt}

and for $\sqrt h$:
\begin{equation}
-\frac38\kappa^2\frac{\partial^2\Psi}{\partial{\cal V}^2}
-\frac{21}{8}\kappa^2\rho^{(3)}\frac{\partial\Psi}{\partial{\cal V}}
- \frac34\kappa^2\rho^{(5)}\frac{\partial\Psi}{\partial{\cal R}} +
\frac3{2\kappa}\alpha\frac{\partial\Psi}{\partial{\cal V}} +
\frac2{\kappa^2}\Lambda\Psi=0.\label{L}
\end{equation}

Now we must consider two cases:
\begin{enumerate}
\item $\Psi$ does not depend on ${\cal R}$. Then from equations above 
we find
\begin{equation}
\frac18\kappa\gamma\frac{\partial\Psi}{\partial{\cal V}} +
\frac1{\kappa^2}\Psi=0,\label{R1}
\end{equation}
\begin{equation}
-\frac38\kappa^2\frac{\partial^2\Psi}{\partial{\cal V}^2}
-\left(\frac{21}{8}\kappa^2\rho^{(3)} -
\frac3{2\kappa}\alpha\right)\frac{\partial\Psi}{\partial{\cal V}} +
\frac2{\kappa^2}\Lambda\Psi=0.\label{L1}
\end{equation}
To make equations (\ref{R1}) and (\ref{L1}) consistent with each 
other,
the coefficients must satisfy the relation
\begin{equation}
\frac{2\Lambda}{\kappa^2}\gamma^2 + \left(\frac{21\rho^{(3)}}
{\kappa}-
\frac{3\alpha}{16\kappa^4} \right)\gamma-24\frac{1}{\kappa^4}=0
\label{cond1}
\end{equation}
and the solution is
\begin{equation}
\Psi({\cal V})=\exp\left(-\frac{8}{\kappa^3\gamma}{\cal V}\right).
\label{sol1}
\end{equation}
\item In the case when $\Psi$ depends on both ${\cal V}$ and 
${\cal R}$,
we must take into account the fact that $\Psi=\tilde\Psi({\cal
V})\exp\left(\frac{\gamma}{\kappa}{\cal R}\right)$. Then we find 
the
following equations
\begin{equation}
-\frac18\kappa\gamma\frac{\partial\tilde\Psi}{\partial{\cal V}}+
\left(\frac78\kappa\gamma\rho^{(3)} +
\frac1{2\kappa^2}\alpha\gamma +
\frac1{\kappa^2}\right)\tilde\Psi=0,\label{R3}
\end{equation}
\begin{equation}
-\frac38\kappa^2\frac{\partial^2\tilde\Psi}{\partial{\cal V}^2}
-\left(\frac{21}{8}\kappa^2\rho^{(3)}-
\frac3{2\kappa}\alpha\right)\frac{\partial\tilde\Psi}{\partial
{\cal V}} -
\left(\frac34\rho^{(5)}\frac{\gamma}{\kappa}-
\frac2{\kappa^2}\Lambda\right)\tilde\Psi=0.\label{L3}
\end{equation}
From equation (\ref{R3}) we find the solution for $\tilde\Psi$; thus
\begin{equation}
\Psi({\cal V},{\cal R})=\exp\left\{\left(
7\rho^{(3)}+\frac{4\alpha}{\kappa^3}+\frac{8}{\kappa^3\gamma}\right)
{\cal V}\right\}
\exp\left\{\frac{\gamma}{\kappa}{\cal R}\right\}.\label{sol2}
\end{equation}
Substituting (\ref{sol2}) into (\ref{L3}) we find another condition on
the coefficients $\gamma$ and $\alpha$. Together with (\ref{cond1}) it
forms a system of equations which turns into a sixth order algebraic
equation for
$\gamma$. Each of the solutions of the latter defines unambigously the
operator and the set of its solutions.
\end{enumerate}

\section{Quantum potential interpretation}

In the previous section we found a class of solutions of the Wheeler 
-
De Witt equation. However the physical interpretation of these
"wavefunctions of the universe" is quite obscured.

It turns out that there exists a nice device which makes it possible 
to
interpret a wavefunction in terms of modified particle or field
dynamics. This approach is an extension of works of David Bohm on
interpretation of quantum mechanics \cite{Bohm} and was presented in
\cite{qp} (see also \cite{shtanov}.) It should be stressed at this 
point
that we use here the quantum potential approach solely as a technical
device
to picture the wavefunction  and we do not attempt to discuss the 
issue
of interpretation of quantum theory.

As compared with the work \cite{qp} here we have to do  with one
important modification resulting from the presence of the $L$ term
in our Wheeler - De Witt operator. Therefore we repeat firs the most
important steps, referring the reader to the original paper
\cite{qp} for more details.

Assume that the wavefunction of the universe is of the form
\begin{equation}
\Psi=e^{\Gamma}e^{i\Sigma}.
\end{equation}
The idea is to substitute this wavefunction to the Wheeler - De Witt
equation and consider only the real part of the resulting equation.
We obtain
$$
-\kappa^2G_{abcd}(x)\frac{\delta\Sigma}{\delta
h_{ab}(x)}\frac{\delta\Sigma}{\delta 
h_{cd}(x)}+\frac1{\kappa}\sqrt{h(x)}
(R(x)+2\Lambda)
+\Re(L)_{ab}(x)\frac{\delta\Gamma}{\delta h_{ab}(x)}
$$
\begin{equation}
-\Im(L)_{ab}(x)\frac{\delta\Sigma}{\delta h_{ab}(x)}
+e^{-\Gamma}
\kappa^2\left(\frac{\delta^{2}e^{\Gamma}}{\delta 
h^{2}}\right)_{ren}(x)=0,
\label{HJ}
\end{equation}
where $\Re(L)_{ab}$ and $\Im(L)_{ab}$ denote the real and imaginary 
part
of $L_{ab}$, respectively
In the last term we used the abbreviated notation to indicate that
the action of the second functional derivative is renormalised.
Now one identifies momenta with the (functional) gradient of 
$\Sigma$,
to wit
\begin{equation}
p^{ab}(x) = \frac{\delta\Sigma}{\delta h_{ab}(x)}.\label{p}
\end{equation}
Then the first two terms in (\ref{HJ}) are identical with the 
hamiltonian
constraint of classical general relativity. The remaining terms 
are
understood as quantum corrections (if we reintroduced $\hbar$ all 
these
terms would become multiplied by $\hbar^2$.)
The wave function is subject to the second set of equations, namely 
the
ones enforcing the three dimensional diffeomorphism invariance. 
These
equations read (for imaginary part)
\begin{equation}
\nabla^a\frac{\delta\Sigma}{\delta h_{ab}(x)} = \nabla^a\, p_{ab} 
=0\label{3diff}
\end{equation}
Thus our theory is defined by two equations (\ref{HJ}) with 
functional
derivatives of $\Sigma$ replaced by $p^{ab}$ as in (\ref{p}), and
(\ref{3diff}). Now we can follow without any alternations the 
derivation
of Gerlach \cite{gerlach} to obtain the full set of ten equations
governing the quantum gravity theory in quantum potential approach
\begin{eqnarray}
0&=& {\cal H}^a = \nabla_a\, p^{ab} ,\label{diff1}\\
0&=& {\cal H}_{\bot} =
-\kappa^2G_{abcd}(x)p^{ab}p^{cd}+\frac1{\kappa^2}\sqrt{h(x)}
(R(x)+2\Lambda)\nonumber \\
&+&\Re(L)_{ab}(x)\frac{\delta\Gamma}{\delta h_{ab}(x)}
-\Im(L)_{ab}(x)p^{ab}
+\kappa^2e^{-\Gamma}
\left(\frac{\delta^{2}\,e^{\Gamma}}{\delta h^{2}}\right)_{ren}(x),
\label{HQ}\\
&&\dot{h}_{ab}(x,t) = \left\{ h_{ab}(x,t),\, {\cal 
H}[N,\vec{N}]\right\},\label{1}\\
&&\dot{p}^{ab}(x,t) = \left\{ p^{ab}(x,t),\, {\cal 
H}[N,\vec{N}]\right\}.\label{2}
\end{eqnarray}
In equations above, $\{\star,\, \star\}$ is the usual Poisson 
bracket, and
\begin{equation}
{\cal H}[N,\vec{N}]=\int\, d^3x \left( N(x){\cal H}_{\bot}(x) + 
N^a(x){\cal
H}_a(x)\right)
\end{equation}
is the total hamiltonian (which is a combination of constraints). It
might seem puzzling at the first sight why to a single wavefunction
there corresponds a set of equations with, clearly, many solutions. 
The
resolution of this problem is that the wavefunction, as a rule, is
sensitive only to some aspects of the configuration. For example, the
wavefunction (\ref{sol1}) depends on ${\cal V}$ only, and thus any
configuration with given volume leads to the same numerical value of 
it.
The above dynamical equations provide us with much more detailed
information concerning the dynamics of the system than the 
wavefunction
alone. 
\newline

Now we apply this formalism to the case of the wavefunction depending
only on ${\cal V}$, (\ref{sol1}) Then $\Gamma =
-\frac{8}{\kappa^3\gamma\gamma^*}{\Re(\gamma)\cal V}$.
Let us inspect equation (\ref{HQ}). The term
$$
\left(\frac{\delta^{2}e^{\Gamma}}{\delta h^{2}}\right)_{ren}(x)
$$
provides us only with modification of the cosmological constant. The 
term
$$
\Re(L)_{ab}(x)\frac{\delta\Gamma}{\delta h_{ab}(x)}
$$
modifies both the cosmological constant {\em and} the coefficient of 
the
term $\sqrt h R$. Taken together these modification can be written as
$$
\frac1{\kappa^2}\sqrt h(R+2\Lambda)\;
\Longrightarrow\;
\frac1{\tilde\kappa^2}\sqrt h(R+2\tilde\Lambda),
$$
where $\tilde\kappa$ and $\tilde\Lambda$ are modified gravitational 
and
cosmological constants, respectively. Our final, modified, 
hamiltonian
constraint reads therefore
$$
{\cal H}_\bot(x) = -
\kappa^2G_{abcd}(x)p^{ab}p^{cd}+\frac{1}{\tilde\kappa^2}\sqrt{h(x)}
(R(x)+2\tilde\Lambda)
$$
\begin{equation}
-\frac{\Im(\alpha)}{\kappa}h_{ab}p^{ab} -
\kappa\Im(\gamma)\left(R_{ab}p^{ab}- \frac14 h_{ab}p^{ab}R\right).
\end{equation}

Now it is clear that the modified hamiltonian above is {\em not} a 
first
class constraint. This follows from the presence of the terms linear 
in
$p$. Thus the effective theory has less symmetries than the classical
general relativity. It follows then that the parameter $N$ in the
definition of total hamiltonian is not free anymore, rather it should 
be
fixed by the requirement that the hamiltonian is time independent, to
wit,
\begin{equation}
\left\{{\cal H}_\bot(x), {\cal H}[N, \vec{N}]
\right\} =0 \;\;\; \mbox{(weakly).}
\end{equation}
It should also be noted that even if the terms linear in $p$ were not
present (if the Poisson bracket constraint algebra would close), we 
still
would not be able to recover the standard four dimensional
action $\int d^4x\sqrt g({}^{(4)}R+2\Lambda)$ from the hamiltonian 
action
$$
\int d^3xdt\left(p^{ab}\dot h_{ab} - {\cal H}[N,\vec{N}]\right).
$$
The reason is that the coefficients $\kappa$ and $\tilde\kappa$ are 
not
identical and therefore the three curvature and external curvature
components of the four dimensional curvature scalar would be 
multiplied
by different coefficients.

It should be stressed that the situation described for the case of 
this
particular solution is quite generic. The conclusion, we draw from 
these
computation is that the four dimensional general coordinate 
invariance
seems to be broken by quantum corrections (besides, this provides the 
ultimate
solution of the celebrated problem of time.) This symmetry is 
restored
when quantum effects are neglegible (because the ordering problems
leading to introducing the $L$-term disappears in the semiclassical 
limit.)
\newline

\section{Conclusions}

The main problem, we address in this paper was to find a set of
conditions which would make it possible to construct the regularized
Wheeler--De Witt operator and a class of states for which the algebra 
of
quantum constraints is anomaly free. It turned out that this space of 
states
is quite
modest but we were able to find some physical states (solutions of
quantum gravity.) We then tried to find an interpretation of one of 
the
solutions employing the method of quantum potential. We found that 
the
resulting modified 3+1 theory does not possess the symmetry of time
translation anymore.
It is hard to say at this moment if this is just an artefact of
employing the
canonical quantisation method, where, by construction of the 
formalism,
the time translation symmetry is
very volnurable from the very beginning or if it signifies some real
physical effect at the Planck scale. It may also happen to be an
artefact of the quantum potential interpretation of the wavefunction.
These questions should be certainly
further investigated.

Another direction of research is to include the coupling of gravity 
to
matter fields like the scalar field or the supergravity. We are now 
in
position to construct relevant regularised operators in both cases, 
but
it turns out that to find solutions in these cases is (surprisingly?)
hard.

These open problems are subject of intensive investigations and we 
hope
to be able to present the results soon.

\end{document}